\documentstyle[twoside,fleqn,npb,epsfig]{article}
\newcommand{\AmS}{{\protect\the\textfont2
  A\kern-.1667em\lower.5ex\hbox{M}\kern-.125emS}}

\newcommand{\W}{\mbox{$W$}}

\newcommand{\qsq}{\mbox{$Q^2$}}

\newcommand{\rh}{\mbox{$\varrho$}}

\newcommand{\rzqzz}{\mbox{$r_{00}^{04}$}}

\newcommand{\rcuu}{\mbox{$r_{11}^{5}$}}
\newcommand{\rczz}{\mbox{$r_{00}^{5}$}}

\newcommand{\cosths}{\mbox{$\cos\theta$}}

\newcommand{\gev}{\mbox{\rm GeV}}

\newcommand{\gevsq}{\mbox{${\rm GeV}^2$}}

\newcommand{\nbinv}{\mbox{${\rm nb^{-1}}$}}
\newcommand{\pbinv}{\mbox{${\rm pb^{-1}}$}}

\newcommand{\bce}{\begin{center}}
\newcommand{\ece}{\end{center}}
\newcommand{\beq}{\begin{equation}}
\newcommand{\eeq}{\end{equation}}
\def\lsim{\mathrel{\rlap{\lower4pt\hbox{\hskip1pt$\sim$}}
    \raise1pt\hbox{$<$}}}         
\def\gsim{\mathrel{\rlap{\lower4pt\hbox{\hskip1pt$\sim$}}
    \raise1pt\hbox{$>$}}}         
\def\ar#1#2#3   {{\em Ann. Rev. Nucl. Part. Sci.} {\bf#1} (#2) #3}
\def\err#1#2#3  {{\it Erratum} {\bf#1} (#2) #3}
\def\ib#1#2#3   {{\it ibid.} {\bf#1} (#2) #3}
\def\ijmp#1#2#3 {{\em Int. J. Mod. Phys.} {\bf#1} (#2) #3}
\def\jetp#1#2#3 {{\em JETP Lett.} {\bf#1} (#2) #3}
\def\mpl#1#2#3  {{\em Mod. Phys. Lett.} {\bf#1} (#2) #3}
\def\nim#1#2#3  {{\em Nucl. Instr. Meth.} {\bf#1} (#2) #3}
\def\nc#1#2#3   {{\em Nuovo Cim.} {\bf#1} (#2) #3}
\def\np#1#2#3   {{\em Nucl. Phys.} {\bf#1} (#2) #3}
\def\pl#1#2#3   {{\em Phys. Lett.} {\bf#1} (#2) #3}
\def\prep#1#2#3 {{\em Phys. Rep.} {\bf#1} (#2) #3}
\def\prev#1#2#3 {{\em Phys. Rev.} {\bf#1} (#2) #3}
\def\prl#1#2#3  {{\em Phys. Rev. Lett.} {\bf#1} (#2) #3}
\def\ptp#1#2#3  {{\em Prog. Th. Phys.} {\bf#1} (#2) #3}
\def\rmp#1#2#3  {{\em Rev. Mod. Phys.} {\bf#1} (#2) #3}
\def\rpp#1#2#3  {{\em Rep. Prog. Phys.} {\bf#1} (#2) #3}
\def\sjnp#1#2#3 {{\em Sov. J. Nucl. Phys.} {\bf#1} (#2) #3}
\def\spj#1#2#3  {{\em Sov. Phys. JEPT} {\bf#1} (#2) #3}
\def\zp#1#2#3   {{\em Zeit. Phys.} {\bf#1} (#2) #3}
\title{Elastic $\rho$ meson production at HERA}

\author{B. Clerbaux \address{Universit\'e Libre de Bruxelles, CP 230, Bd du Triomphe, 
B-1050 Brussels, Belgium. \\ \ \ e-mail: clerbaux@hep.iihe.ac.be} \\
\vspace*{0.2cm} 
For the H1 Collaboration }

\begin{document}

\begin{abstract}
Results on elastic electroproduction of $\rho$ mesons are presented
for a photon virtuality in the range $1 < Q^2 < 60~{\rm GeV^2}$ and 
for a hadronic centre of mass energy in the range $30 < W < 140$~{\rm GeV}.
The shape of the ($\pi \pi$) mass distribution is discussed and 
measurements of the cross section dependences on $Q^2$, $W$ and $t$ (the four-momentum 
transfer squared to the proton) are presented. The full set of \rh\ spin density matrix 
elements is measured, providing information on the \rh\ meson and on the photon
polarisation states. In particular, the ratio $R$ of longitudinal to transverse 
$\gamma^* p$ cross section is determined.
\end{abstract}

\maketitle

\section{Introduction}

We present results~\cite{barbara} on elastic electroproduction of \rh\ mesons:
$ e+p \rightarrow e+p+\rh $, the \rh\ meson decaying into two pions
($\rho \rightarrow \pi^+ \pi^-$, BR $\simeq$ 100 \%). The data were 
collected in 1995
and 1996 by the H1 detector, corresponding respectively to an integrated luminosity 
of 125 \nbinv\ and 3.87 \pbinv. The kinematical range covered in \qsq,
\W\ and $t$ is the following:
$1 < \qsq < 60$ \gevsq, $30 < W < 140$  GeV and $|t| < 0.5$ \gevsq.
The \rh\ meson having the same quantum numbers as the
photon ($J^{PC} = 1^{--}$), the $\gamma^* p$ interaction
is mediated by the exchange of a colourless
object, called the pomeron in the Regge model. It is important
to understand the pomeron in terms of partons in
the framework of the QCD theory.

\section{Models}

Quantitative predictions in perturbative QCD are possible when
a hard scale is present in the interaction. For \rh\ meson production,
this scale can be given by \qsq\ (\qsq\ $\gsim$ several \gevsq).
Most models rely on the fact that, at high energy
in the proton rest frame, the photon fluctuates into a
$q\bar{q}$ pair a long time before the interaction, and recombines
into a \rh\ meson a long time after the interaction.
The amplitude ${\cal M}$ then
factorizes in three terms:
${\cal M} \propto \psi_{\lambda_{\rho}}^{\rho \ *}
 \  T_{\lambda_{\rho} \lambda_{\gamma}} \  \psi^\gamma_{\lambda_{\gamma}}$
where  $T_{\lambda_{\rho} \lambda_{\gamma}}$ are the interaction helicity amplitudes
($\lambda_\gamma$ and $\lambda_\rho$ being the helicities of the
photon and the \rh\ meson, respectively) and $\psi$ represent the wave functions.
In most models, the $q\bar{q}-p$ interaction is described by 2 gluon
exchange. The cross section is then proportional to the square of the
gluon density in the proton:
$ \sigma_{\gamma p} \sim  \alpha_s^2(Q^2) / Q^6 \cdot \left
| xg(x, Q^2) \right| ^2  \label{eq:gluon} $.
The main uncertainties of the models come from the
choice of the scale, of the gluon distribution
parametrisation, of the \rh\ meson wave function (Fermi motion), and from the neglect of
off-diagonal gluon distributions and of higher order corrections.

\section{Signal}
The shape of the ($\pi\pi$) mass distribution has been studied as a function of
\qsq. The mass distributions are skewed compared to a relativistic Breit-Wigner profile:
enhancement is observed in the low mass region and suppression in the
high mass side. This effect has been attributed to an interference between 
the resonant and the non-resonant production of two pions.
The skewing of the mass distribution is observed to decrease with \qsq.
%
\begin{figure}[htb]
  \setlength{\unitlength}{1cm}
  \begin{picture}(7.0,11.5)
  \put(0.0,7.6){\epsfig{file=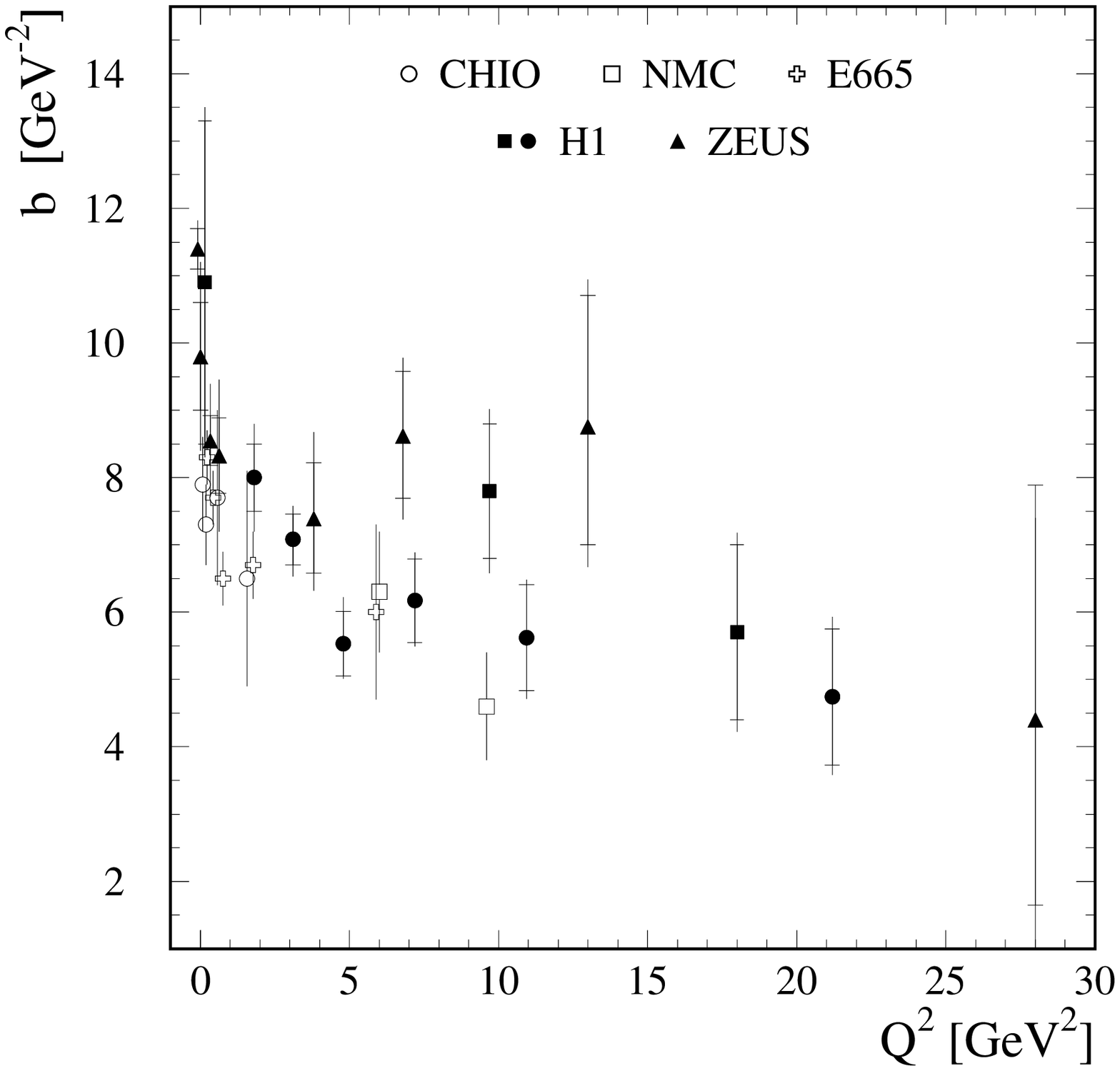,width=7cm,height=4.2cm}}
  \put(0.1,3.25){\epsfig{file=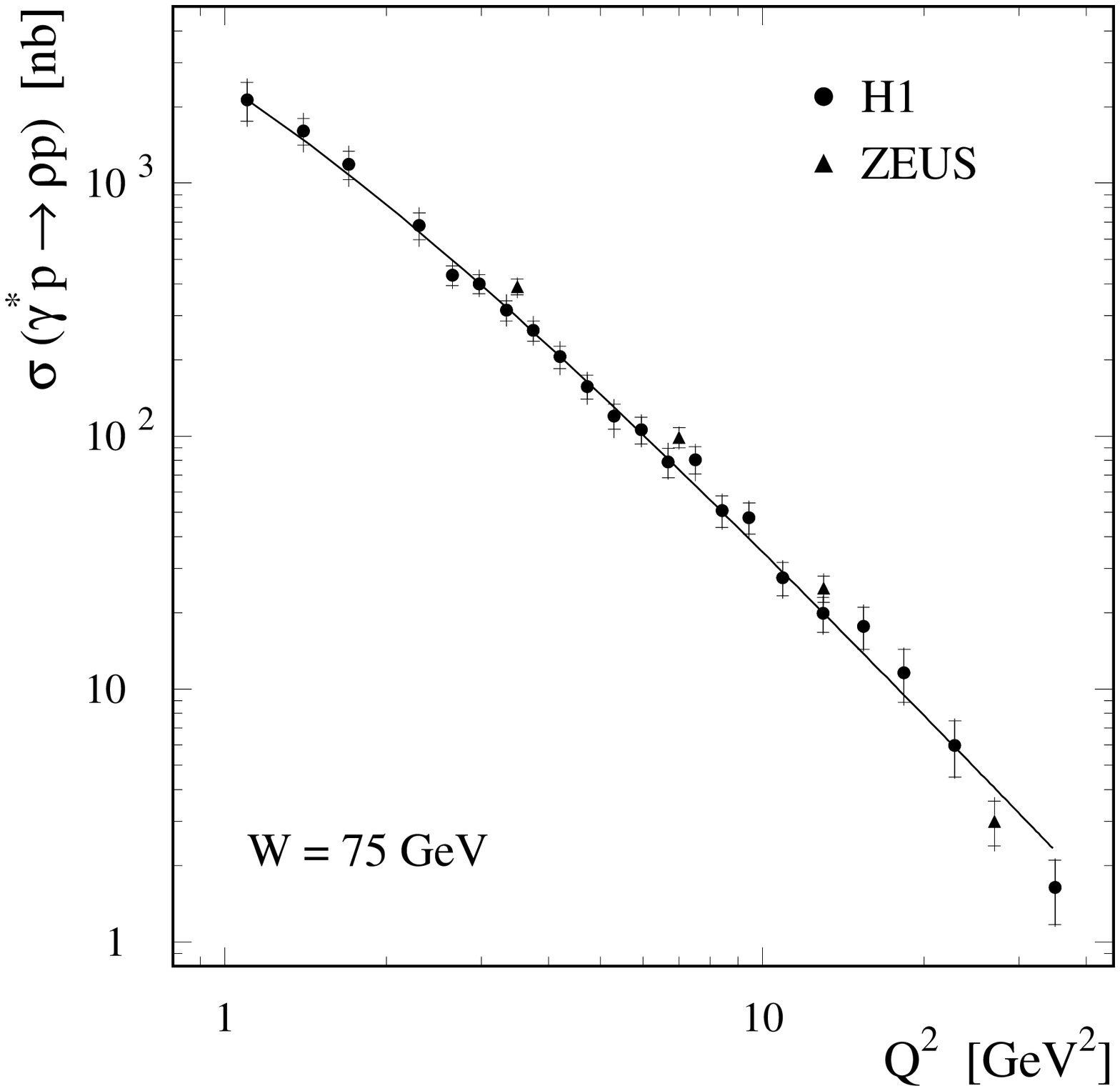,width=7cm,height=4.2cm}}
  \put(0.2,-1.5){\epsfig{file=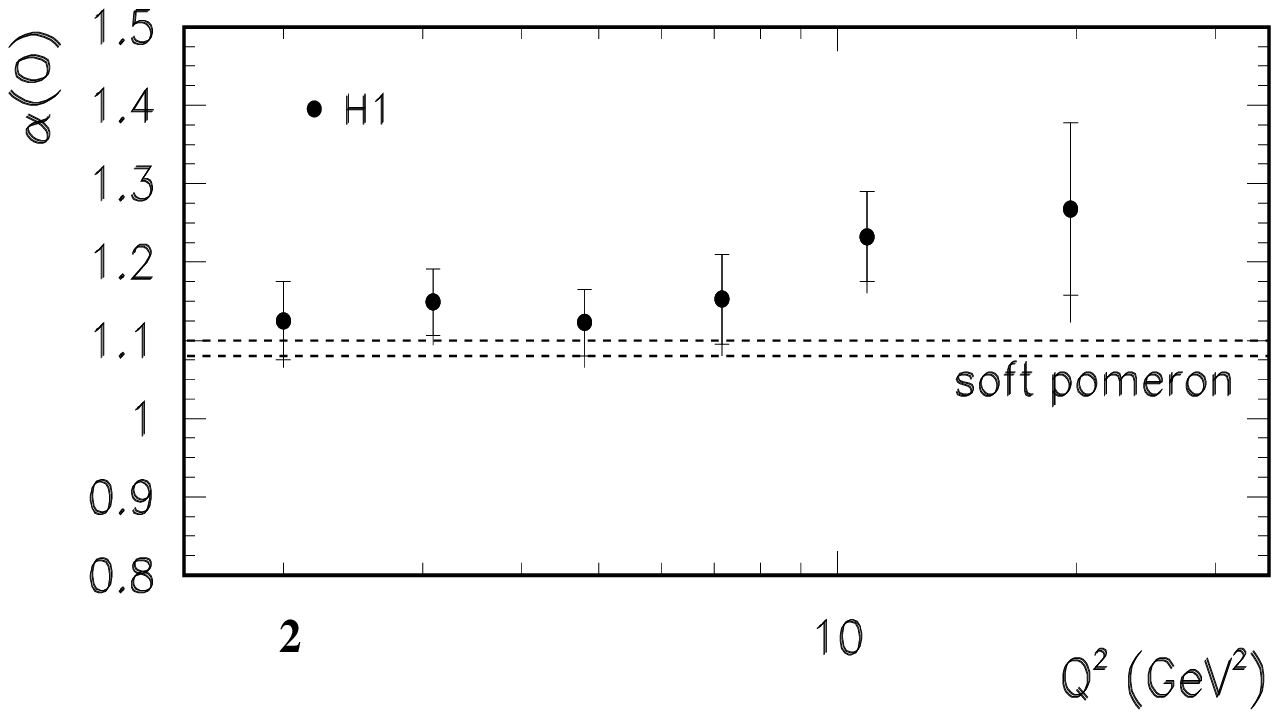,width=6.4cm,height=4.5cm}}
  \put(3.5,7.45){(a)}
  \put(3.5,3.2){(b)}
  \put(3.5,-1.1){(c)}
  \end{picture}
  \vspace*{0.3cm}
  \caption{ (a) $b$ slope parameter as a function of \qsq;
          (b) $\sigma (\gamma^* p \rightarrow \rho p)$ 
             as a function of \qsq\ for $W = 75~\gev$ (full line: see text);
          (c) the intercept $\alpha(0)$ as a function of \qsq.}
\label{fig:cross}
\vspace*{-0.5cm}
\end{figure}
\section{Cross sections}

$\bullet$ $t$ dependence: the data present the characteristic exponential 
falling off of the $t$ distribution $\sigma \propto {\rm exp} (-b |t|)$. 
The $b$ slope parameter, measured for different \qsq\ interval, 
is shown in Fig~\ref{fig:cross}a,
confirming the decrease of $b$ when \qsq\ increases from
photoproduction to the deep-inelastic domain,
reflecting the decrease of the transverse size of the virtual photon.

$\bullet$ \qsq\ dependence: Fig.~\ref{fig:cross}b presents the \qsq\ dependence
of the $\sigma (\gamma^* p \rightarrow \rho p)$ cross section for \W\ = 75 GeV.
The data are well described by the parametrisation (\qsq\ + $m_\rho^2$)$^n$, with
$n$ = 2.24 $\pm$ 0.09 (full line).

$\bullet$ \W\ dependence: the \W\ dependence of 
$\sigma (\gamma^* p \rightarrow \rho p)$ was measured
for different \qsq\ values, and
the parametrisation $\sigma \propto W^{\delta}$ was fitted to the data.
In a Regge context, $\delta$ can be related to the exchange
trajectory~\cite{barbara} and
the values of the intercept $\alpha(0)$ are shown in Fig.~\ref{fig:cross}c.
The measurements are compared to the values $1.08 - 1.10$ obtained from fits to
the total and elastic hadron--hadron cross sections.
They suggest that the intercept of the effective trajectory
governing high \qsq\ \rh\ electroproduction is larger than that describing
elastic and total hadronic cross sections.
The strong rise of the cross section with \W, 
observed at high \qsq, is in agreement with perturbative 
QCD prediction $ \sigma_{\gamma p} \sim  | xg(x, Q^2) | ^2$.

\section{Helicities studies}

\begin{center}
\begin{figure}[tb]
\setlength{\unitlength}{1.0cm}
\begin{picture}(7.0,15.0)
\put(-0.5,7.0){\epsfig{file=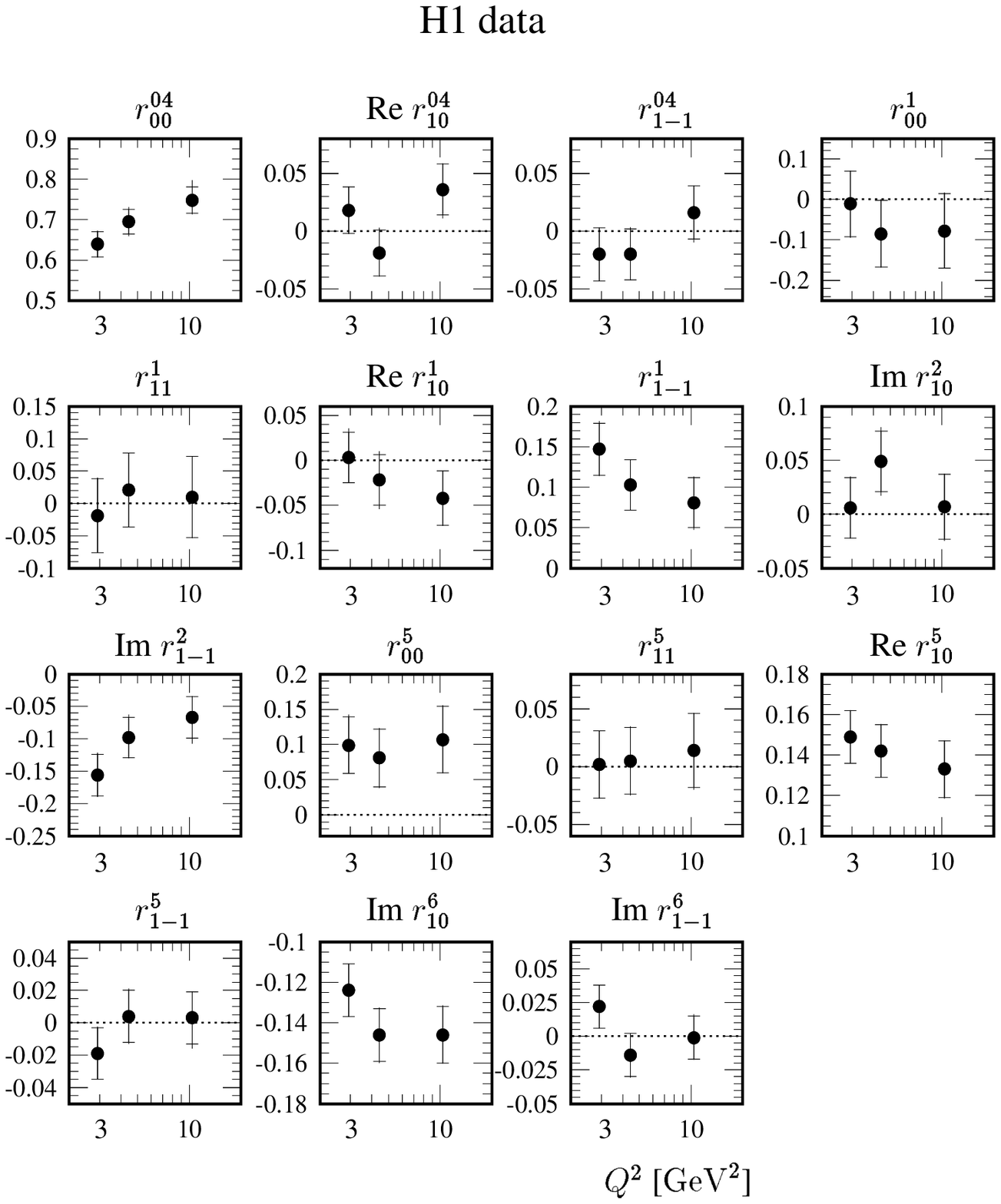,%
            bbllx=55pt,bblly=189pt,bburx=450pt,bbury=693pt,%
            height=8.2cm,width=6.8cm}}
\put(0.2,4.0){\epsfig{file=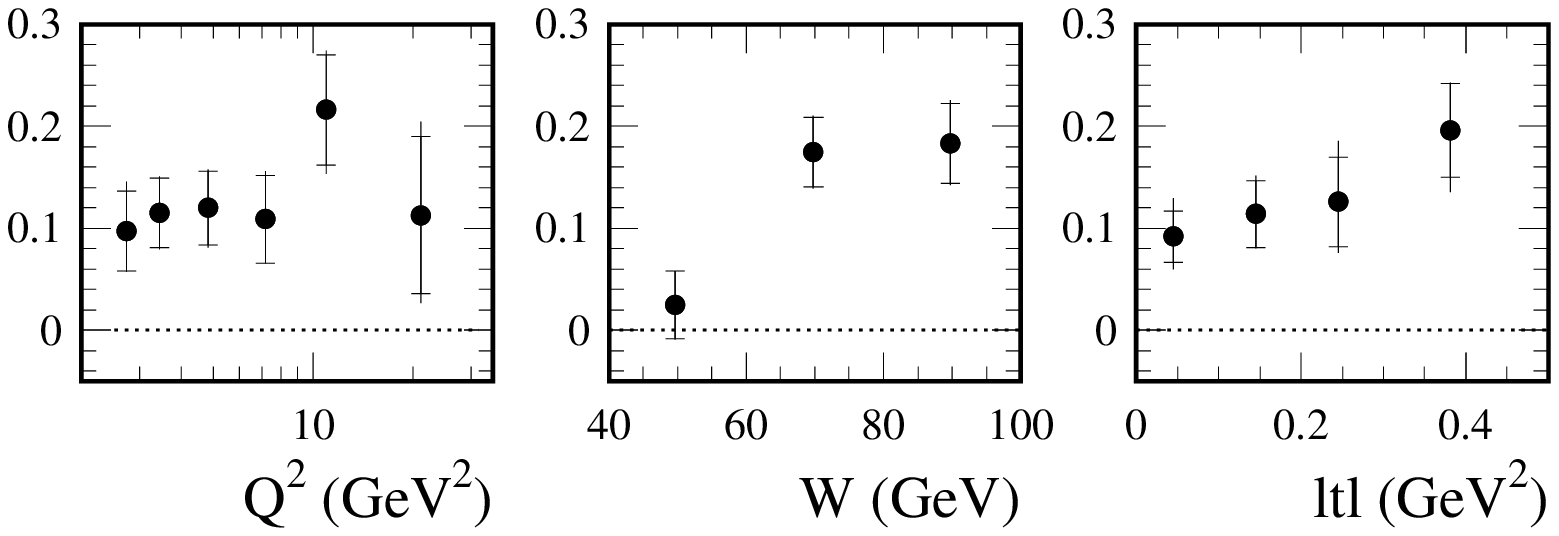,%
            bbllx=70pt,bblly=487pt,bburx=536pt,bbury=658pt,%
            height=3.0cm,width=7.cm}}
\put(0.,-1.0){\epsfig{file=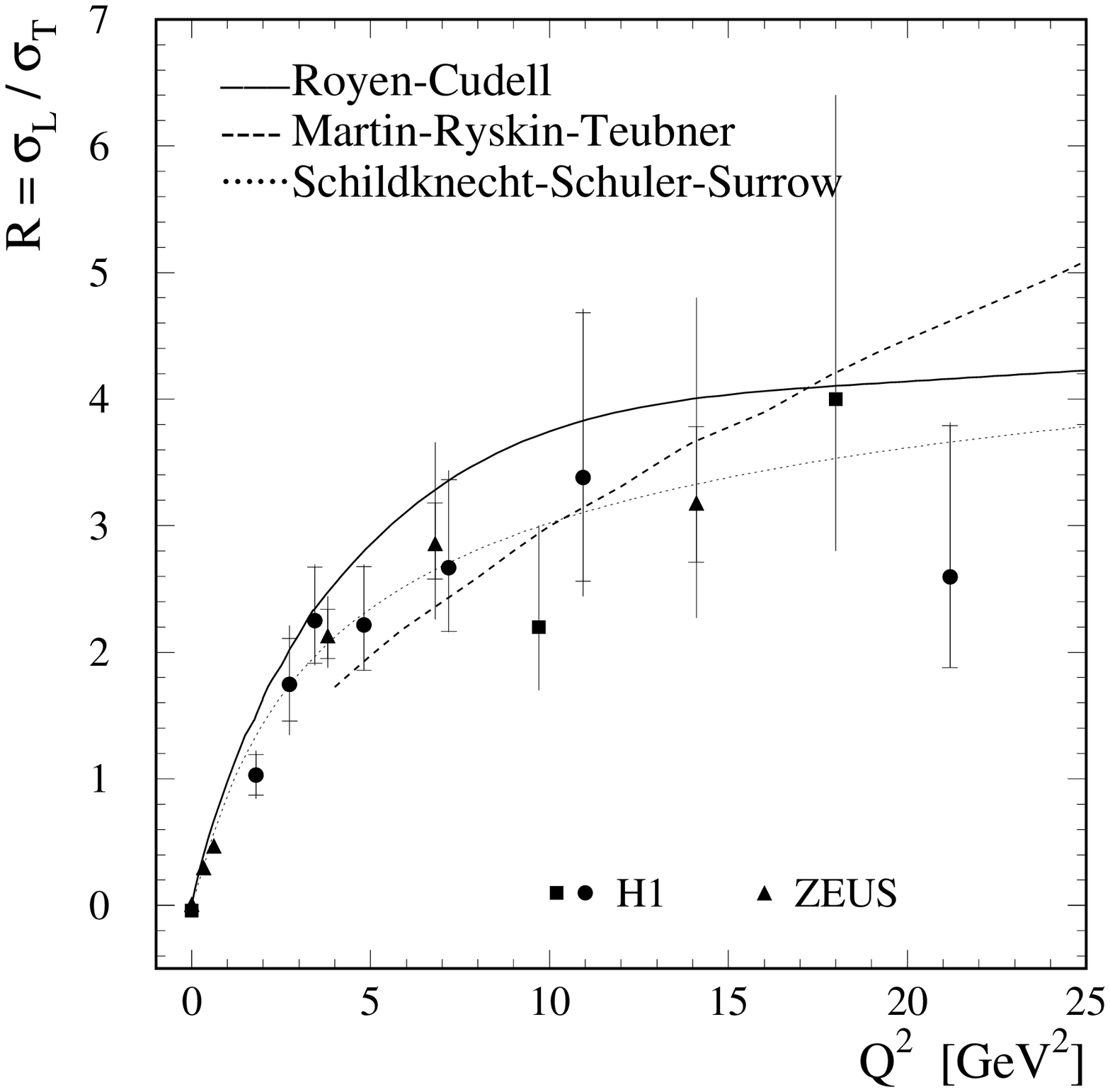,%
         height=5.0cm,width=7.cm}}
\put(6.2,8.6){(a)}
\put(5.8,7.5){ $\downarrow$ (b)}
\put(5.5,3.0){(c)}
\put(0.5,7.1){ $(2 \rcuu + \rczz)$}
\end{picture}
\caption{(a) Spin density matrix elements measured for three values of \qsq.
  (b) Measurements of the combination of matrix elements $2 \rcuu + \rczz$,
  as a function of \qsq, \W\ and $|t|$, obtained from fits to the $\Phi$
  distributions. In both figures, the dashed lines indicate the null values 
  expected in the case of SCHC.
  (c) The ratio $R$ of the longitudinal to transverse photon cross sections 
  as a function of \qsq. The curves are the predictions of the models (see text).}
\label{fig:matele}
\end{figure}
\end{center}

The study of the angular distributions of the production and decay of the
\rh~meson gives information on the photon and \rh\ polarisation states.
In the helicity frame, three angles are used: the polar
($\theta$) and azimuthal ($\varphi$) angles of the $\pi^+$ direction
in the \rh\ meson centre of mass system (cms), and the
$\Phi$ angle between the electron scattering plane and the
\rh\ meson production plane, in the hadronic cms. The decay angular
distribution $W(\cosths, \varphi, \Phi)$ is a function of
15 matrix elements $r^{\alpha}_{ij}$
and $r^{\alpha \beta}_{ij}$, which are related to the
helicity amplitudes $T_{\lambda_{\rho} \lambda_{\gamma}}$. 
Figure~\ref{fig:matele}a presents the measurement
of the 15 matrix elements (using the ``moment method")
as a function of \qsq. In case of s-channel
helicity conservation (SCHC), the helicity of the vector meson is the same
as that of the photon 
($T_{\lambda_\rho \lambda_\gamma} = T_{01} = T_{10} = T_{1-1} = T_{-11} = 0$),
and 10 of the matrix elements vanish (dotted lines in 
Figs.~\ref{fig:matele}a and~\ref{fig:matele}b). The measurement of
the matrix elements are in agreement which SCHC except for the \rczz\ element,
which is observed to be significantly different from zero. This element is
proportional to the single helicity flip amplitude 
$T_{\lambda_\rho \lambda_\gamma} = T_{01}$.

Another way to
extract the \rczz\ matrix element is to study the $\Phi$ distribution. Indeed
the decay angular distribution $W(\Phi)$ depends on the combinaison 
$(2\rcuu+ \rczz)$. Figure~\ref{fig:matele}b presents the result of the 
fits in different \qsq, \W\ and $|t|$ bins.
Again, we observe a clear deviation of the \rczz\ parameter from the null
value expected for SCHC.  The ratio of helicity flip to 
non helicity flip amplitudes is hence estimated to be $8.0 \pm 3.0$ \%.

The ratio of the longitudinal to the transverse cross
section, $R = \sigma_L / \sigma_T$, can be extracted using the measurement of
the \rzqzz\ matrix element. $R$ is observed to
increase with \qsq, and to reach the value
$R$ = 3 -- 4 for \qsq\ $\simeq$ 20 \gevsq\ (see Fig.~\ref{fig:matele}c).
The \qsq\ dependence of the ratio $R$ is well described by the perturbative QCD models of
Royen and Cudell~\cite{royen} (full line), and of Martin, Ryskin and
Teubner~\cite{mrt} (dashed line) and by the model of Schildknect, 
Schuler and Surrow~\cite{sss} (dotted line)
based on generalised vector dominance model (GVDM). 
The following hierarchy between the helicity amplitudes, observed in the data:
$|T_{00}| > |T_{11}| > |T_{01}| > |T_{10}|, |T_{1-1}|$, is in agreement
with perturbative QCD calculations performed by Ivanov and Kirschner~\cite{ivanov}.

\vspace{-0.7cm}
\section{Conclusions}
The elastic electroproduction of \rh\ mesons has been studied at HERA
with the H1 detector in a wide kinematical domain: 
$1~< Q^2 < 60$~GeV$^2$ and 30 $<$ \W\ $<$ 140~GeV.
Measurements of the cross section
$\sigma ( \gamma^* p \rightarrow \rho p )$
show an indication for an increasingly strong energy dependence when
\qsq\ increases. 
Full helicity studies have been performed showing a small but
significant violation of SCHC. The \qsq\ dependence of the 
ratio $R = \sigma_L / \sigma_T$ was measured and is 
well described by two models based on perturbative QCD 
~\cite{royen,mrt} and by a model based on GVDM~\cite{sss}.

\end{document}